\documentclass[12pt]{JHEP3}
\usepackage{graphicx}
\usepackage{amsfonts, amsmath}
\usepackage{mathrsfs}
\usepackage{setspace}
\usepackage[utf8]{inputenc}
\usepackage{comment}
\usepackage{cite}

\newcommand{\be}{\begin{equation}}
\newcommand{\ee}{\end{equation}}

\title{Constraint structure of the three dimensional massive gravity}

\author{M. Sadegh\footnote{m.sadegh@ph.iut.ac.ir},
  A. Shirzad\footnote{shirzad@ipm.ir},
\\
{\it Department of Physics, Isfahan University of Technology \\
P.O.Box 84156-83111, Isfahan, IRAN, \\
School of Physics, Institute for Research in Fundamental Sciences (IPM)\\
P.O.Box 19395-5531, Tehran, IRAN.} }

\abstract{Constraint analysis of the three-dimensional massive gravity, the so-called new massive gravity, is
studied in the Palatini formalism. We show that amongst 6 components of the metric, 2 are dynamical,
which is compatible with the existence of one vector massive graviton in the linearized theory
(Fierz-Pauli theory).}


\preprint{\today}

\begin{document}


\section{Introduction}
It is well known that theories with general covariance are
constrained systems \cite{refDirac1,ref
Dirac2,refHRT,refBergman1}. In other words, their equations of
motion in Lagrangian formalism lead to acceleration-free
relations. On the other hand, constructing the Hamiltonian
formulation for such theories needs care in order to consider
the constraints. The {\textit{primary}} constraints emerge in the phase
space whenever the momenta are not independent functions
of velocities. The {\textit{secondary}} constraints come out as the
result of the consistency of primary constraints.

The most difficulty in Hamiltonian treatment of general
covariant theories is that the action depends on the second
derivatives of the metric, as well. In this situation a wellbehaved
Hamiltonian system is not recognized, or at least is
not agreed upon, even when a system is not constrained.
However, for Einstein-Hilbert gravity
\cite{refSundermeyer} or, for instance, Ho\v{r}ava gravity
\cite{refhorava} one may use one may use the Arnowitt-Deser-Misner
variables\cite{refADM} whose advantage is that the Lagrangian does not
contain accelerations when written in terms of these variables.
This may not happen for an arbitrary general covariant
Lagrangian.

 The other possibility is using the so-called Palatini
formalism in which the Christoffel symbols are considered
as independent variable. For Einstein-Hilbert gravity, this
approach does work well. The reason is the relation between
the Christoffel symbols and the derivatives of the
metric results naturally from the equation of motion of
Christoffel symbols. It was shown recently
\cite{ref2} that for a
gravitation theory of the Lovelock-type, the Palatini formalism
is fine. However, for an arbitrary model the equations
of motion give no guaranty about the relations of
Christoffel symbols and derivatives of the metric.
Therefore, one needs to add them to the Lagrangian using
Lagrange multipliers, which should be considered as additional
variables in the Lagrangian formalism.

The three-dimensional gravity has attracted intensive
interest in recent years. One reason is that it is possible
to construct nontrivial renormalizable models in three dimensions.
Among so many attractive features, investigating
the Hamiltonian structure of the models is noticeable.
The topological massive gravity
\cite{refTMG}, which is generally covariant on a closed manifold, is
one of the most important ones. The Hamiltonian structure of TMG
is discussed in some papers\cite{refcgmb}.

Recently Begshoeff, Hohm and Townsend proposed a model\cite{ref0}
for three dimensional massive gravity (the so called new massive
gravity) which preserves parity and possesses general covariance
on an arbitrary manifold. Linearization around the flat metric of
this model leads to Pauli-Fierz action describing massive
graviton. Then Deser\cite{refdeser} showed that this model is
finite and ghost-free. Oda and Nakasone\cite{refodanakasone}
showed afterward that the model is unitary and renormalizable.
Clem\`{e}nt also gave some black hole solutions of the
model\cite{refclement}.
The Lagrangian of new massive gravity (NMG), at one
hand, includes accelerations (i.e. second order derivatives
of the metric), which make it necessary to use Christoffel
symbols (or combinations of them) as auxiliary fields. On
the other hand, the model is not of the Lovelock-type.
These peculiarities make the Hamiltonian treatment and
constraint structure of NMG much more difficult.
Moreover, the existence of quadratic terms with respect
to $R_{\mu\nu}$ makes it difficult to use the Arnowitt-Deser-Misner
variables. However, in spite of complicated calculations,
the Hamiltonian treatment of the theory can be followed
carefully. This is what we have done in this paper.

Our main task in this work is counting the physical
degrees of freedom of this model. Since the Einstein gravity
in three dimensions has zero degrees of freedom, one
expects roughly that the NMG action, which contains two
more derivatives, should have 2 degrees of freedom; in the
same way as TMG with one more derivative than Einstein-
Hilbert action possesses 1 degree of freedom. This result is
in agreement with the number of massive gravitons in the
linearized model. However, from a theoretical point of
view, it is better to check the validity of such rough arguments
by a careful Hamiltonian analysis.

In Section 2 we introduce the model in the Hamiltonian formalism
and find the primary constraints. In Section 3 we follow the
consistency conditions of the constraints and find the secondary
constraints of system. Section 4 is devoted to our conclusions.

In our work we use Greek indices for space-time components and
Latin indices
 for space components.

\section{Lagrangian and Hamiltonian}
The action of NMG is given as:
\begin{eqnarray}\label{12}
 S=\frac{1}{16\pi G}\int {d^3x \sqrt g\left[\ R-2\Lambda+\frac{1}{m^2} (R_{\mu\nu}
R^{\mu\nu}-\frac{3}{8} R^2)\right]\ },
\end{eqnarray}
where $g$ is the metric determinant, $R_{\mu\nu}$ is the Ricci
tensor and $R$ is the Ricci scalar. We assume that 3d
space-time is torsion-free and the Christoffel symbols
$\Gamma^\lambda_{\mu\nu}$ are symmetric with respect to $\mu$ and
$\nu$. This allows us to introduce new variables
$\xi^\lambda_{\mu\nu}$ via
\begin{eqnarray}\label{11}
 \Gamma^{\lambda}_{\mu\nu}=\xi^{\lambda}_{\mu\nu}-\frac{1}{2} (\delta^{\lambda}_{\mu}\xi^{\sigma}_{\nu\sigma}+
 \delta^{\lambda}_{\nu}\xi^{\sigma}_{\mu\sigma}),
\end{eqnarray}
as in reference \cite{ref1}. The Ricci tensor in terms of
$\xi$ variables contains derivatives in the form of total
3-divergence as
\begin{eqnarray}\label{15-6}
  R_{\mu\nu}=\xi^{\lambda}_{\mu\nu,\lambda}-\xi^{\lambda}_{\mu\sigma}\xi^{\sigma}_{\nu\lambda}+\frac{1}{2}
  \xi^{\sigma}_{\mu\sigma}\xi^{\lambda}_{\nu\lambda}.
\end{eqnarray}
This is, in fact the advantage of using $\xi$'s in comparison with
$\Gamma$'s. In this way, $\xi^0_{\mu\nu}$ are the only variables
whose velocities are present in the Lagrangian.
  As is well known,
in 3 dimensions an action containing higher order derivatives can
not be of Lovelock-type. So the Palatini approach can not be used
without imposing explicitly the relation between the metric
$g_{\mu\nu}$ and auxiliary variables $\xi^\lambda_{\mu\nu}$. Using
the Eq. (\ref{11}) and the definition of Christoffel symbols as
\begin{eqnarray}
\Gamma^\lambda_{\mu\nu}=\frac{1}{2}g^{\lambda\rho}
(g_{\mu\rho,\nu}+g_{\nu\rho,\mu}-g_{\mu\nu,\rho}),
\end{eqnarray}
we have
 \begin{eqnarray}\label{13}
 \Psi_{\alpha\lambda\beta}\equiv g_{\lambda\rho}(\xi^{\rho}_{\alpha\beta}-
\frac{1}{2}(\delta^{\rho}_{\alpha}\xi^{\sigma}_{\beta\sigma}+
 \delta^{\rho}_{\beta}\xi^{\sigma}_{\alpha\sigma}))
 -\frac{1}{2}(g_{\lambda\beta,\alpha}+g_{\lambda\alpha,\beta}-
 g_{\alpha\beta,\lambda})=0.
\end{eqnarray}
Expressions $\Psi_{\alpha\lambda\beta}$ should be considered as
external Lagrangian constraints which should be put by hand in the
Lagrangian with Lagrange multipliers. These Lagrange multipliers
then should be taken into account as new variables in addition to
$g_{\mu\nu}$ and $\xi^\lambda_{\mu\nu}$. In this way, the
following action should be considered instead of the original one
(\ref{12}):
\begin{eqnarray}\label{4}
 S&=&\int d^3x\mathcal{L}=\int d^3x \sqrt {g}[R-2\Lambda+\frac{1}{m^2}(R_{\mu\nu}
R^{\mu\nu}-\frac{3}{8} R^2)]\nonumber \\
  &+&\int d^3x A^{\alpha\lambda\beta}\left(\ g_{\lambda\rho}
  (\xi^{\rho}_{\alpha\beta}-
\frac{1}{2}(\delta^{\rho}_{\alpha}\xi^{\sigma}_{\beta\sigma}+
 \delta^{\rho}_{\beta}\xi^{\sigma}_{\alpha\sigma}))
 -\frac{1}{2}(g_{\lambda\beta,\alpha}+g_{\lambda\alpha,\beta}-
 g_{\alpha\beta,\lambda})\right)\ .
\end{eqnarray}
where $A^{\alpha\lambda\beta}$ is Lagrange multiplier.
Let us enumerate different degrees of freedom before considering
the details of the dynamics of the system. We have six degrees of
freedom $g_{\mu\nu}$ and $3\times 6=18$ auxiliary variables
$\xi^\lambda_{\mu\nu}$, taking into account the $\mu
\leftrightarrow\nu$ symmetry in both cases. In other words, by
adding the auxiliary variables,
number of Lagrangian degrees of freedom are multiplied by 4 to
avoid higher order derivatives (greater than 2) in the equations
of motion. We have also introduced 18 more degrees of freedom
$A^{\alpha\lambda\beta}$, since Eq. (\ref{13}) is symmetric with
respect to indices $\alpha$ and $\beta$ and we have
$A^{\alpha\lambda\beta}=A^{\beta\lambda\alpha}$. Putting all these
points together we have a priory $18+18+6=42$ Lagrangian variables which
is equivalent to $84$ phase space variables.\\
Now we proceed to the Hamiltonian formalism. The canonical momenta
conjugate to $g_{\mu\nu}$, $\xi^\lambda_{\mu\nu}$ and
$A^{\alpha\lambda\beta}$ are defined respectively as
\begin{eqnarray}\label{15}
&&\Pi^{\mu\nu}_{\gamma}=\frac{\partial \mathcal{L}}{\partial \dot
   \xi^{\gamma}_{\mu\nu}}=\sqrt{g}\delta^0_{\gamma}(g^{\mu\nu}+
   \frac{2}{m^2}G^{\mu\alpha\nu\beta}R_{\alpha\beta}),
\end{eqnarray}
\begin{eqnarray}\label{15-1}
  \pi^{\mu\nu}=\frac{\partial \mathcal{L}}{\partial \dot
  g_{\mu\nu}}=-\frac{1}{2}(A^{0\mu\nu}+A^{0\nu\mu}-A^{\mu 0 \nu}),
 \end{eqnarray}
 \begin{eqnarray}\label{15-2}
  P_{\alpha\lambda\beta}=\frac{\partial \mathcal{L}}{\partial \dot
  A^{\alpha\lambda\beta}}=0.
\end{eqnarray}
  where $G^{\mu\alpha\nu\beta}$ is the generalized metric defined
  by
  \begin{eqnarray}
  G^{\mu\alpha\nu\beta}=g^{\mu\alpha}g^{\nu\beta}-\frac{3}{8}g^{\mu\nu}g^{\alpha\beta}.
  \end{eqnarray}
  From eq.(\ref{15}), with $\gamma=0$, we have
  \begin{eqnarray}\label{15-3}
  R_{\mu\nu}=\frac{1}{2}m^2\mathcal{G}_{\mu\alpha\nu\beta}(g^{-1/2}\Pi_0^{\alpha\beta}-g^{\alpha\beta}).
  \end{eqnarray}
  where $\mathcal{G}_{\mu\alpha\nu\beta}$ is the inverse  of the
generalized metric $G^{\mu\alpha\nu\beta}$ such that $\mathcal
{G}_{\mu\alpha\nu\beta} G^{\mu\sigma\nu\gamma}=
\delta_\alpha^\sigma \delta_\beta^\gamma$. Hence, using
(\ref{15-6}) we find
  \begin{eqnarray}\label{15-4}
  \dot\xi^0_{\mu\nu}=\frac{1}{2}m^2g^{-1/2}\mathcal{G}_{\mu\alpha\nu\beta}\Pi_0^{\alpha\beta}+4m^2g_{\mu\nu}
  -\xi^i_{\mu\nu,i}+\xi^{\lambda}_{\mu\sigma}\xi^{\sigma}_{\nu\lambda}-\frac{1}{2}
  \xi^{\sigma}_{\mu\sigma}\xi^{\lambda}_{\nu\lambda}.
  \end{eqnarray}
  Using eqs.(\ref{15-3}) and (\ref{15-4}) the canonical Hamiltonian
density can be derived in the usual way as
$H_C=\int{d^2x\mathcal{H}_C}$, where
\begin{eqnarray}\label{24}
   \mathcal{H}_C&=&\pi^{\mu\nu}\dot g_{\mu\nu}+\Pi_\gamma^{\mu\nu}\dot
   \xi^{\gamma}_{\mu\nu}+ P_{\alpha\lambda\beta}\dot
   A^{\alpha\lambda\beta}-\mathcal{L}\nonumber\\ &=&
   \frac{1}{4}m^2 g^{-1/2}\mathcal{G}_{\mu\alpha\nu\beta}
   \Pi_0^{\alpha\beta}\Pi_0^{\mu\nu}+2\sqrt{g}(\Lambda-3m^2)\nonumber\\
    &+&\frac{1}{2}(A^{i\mu\nu}+A^{i\nu\mu}-A^{\mu i
  \nu})g_{\mu\nu,i}-\Pi_0^{\mu\nu}\left[\ \xi^i_{\mu\nu,i}-
  \xi^{\lambda}_{\mu\sigma}\xi^{\sigma}_{\nu\lambda}+\frac{1}{2}
  \xi^{\sigma}_{\mu\sigma}\xi^{\lambda}_{\nu\lambda} \right]\ \nonumber\\
  &-& A^{\alpha\lambda\beta}g_{\lambda\rho}\left (\ \xi^{\rho}_{\alpha\beta}-
\frac{1}{2}(\delta^{\rho}_{\alpha}\xi^{\sigma}_{\beta\sigma}+
 \delta^{\rho}_{\beta}\xi^{\sigma}_{\alpha\sigma})\right )\  .
   \end{eqnarray}
  In deriving the
Canonical Hamiltonian the following primary constraints resulted
from Eqs. (\ref{15}-\ref{15-2}) are imposed:
   \begin{eqnarray}\label{16}
  &&\phi^{\mu\nu}:=\pi^{\mu\nu}+\frac{1}{2}(A^{0\mu\nu}+A^{0\nu\mu}
  -A^{\mu 0 \nu})\approx 0,\nonumber\\
  &&\Phi_i^{\mu\nu}:=\Pi_i^{\mu\nu}\approx 0 ,\\
  &&\Omega_{\mu\lambda\nu}:=P_{\mu\lambda\nu}\approx
  0, \nonumber
\end{eqnarray}
where the symbol "$\approx$" means weak equality, i.e., equality on
the constraint surface. We recall that primary constraints are
identities amongst coordinates and momenta which follow directly
from the definition of Canonical momenta.
 The total Hamiltonian reads
\begin{eqnarray}
 &&H_T=\int d^3x\mathcal{H}_T\nonumber\\
&&\mathcal{H}_T=\mathcal{H}_C+U_{\mu\nu}\phi^{\mu\nu}
+\lambda^i_{\mu\nu}\Phi_i^{\mu\nu}
+V^{\mu\alpha\nu}\Omega_{\mu\alpha\nu},
\end{eqnarray}
where $U_{\mu\nu}$, $\lambda^i_{\mu\nu}$ and $V^{\mu\alpha\nu}$
are Lagrange multipliers (in the context of Hamiltonian
constrained systems) corresponding to the primary constraints
(\ref{16}) respectively.\\
The fundamental Poisson brackets of field variables are
\begin{eqnarray}
  &&\{g_{\mu\nu}(x),\pi^{\alpha\beta}(y)\}=\Delta^{\alpha\beta}_{\mu\nu}\delta^{(3)}(x-y), \nonumber\\
  &&\{\xi^\lambda_{\mu\nu}(x)
  ,\Pi_\gamma^{\alpha\beta}(y)\}=\delta^\lambda_\gamma
  \Delta^{\alpha\beta}_{\mu\nu}\delta^{(3)}(x-y), \\
  &&\{A^{\alpha\lambda\beta}(x),P_{\mu\gamma\nu}(y)\}=\delta^\lambda_\gamma
  \Delta^{\alpha\beta}_{\mu\nu}\delta^{(3)}(x-y),\nonumber
\end{eqnarray}
where
$\Delta^{\alpha\beta}_{\mu\nu}\equiv\frac{1}{2}(\delta_{\mu}^{\alpha}
\delta_{\nu}^{\beta}+\delta_{\nu}^{\alpha}\delta_{\mu}^{\beta})$.
\section{Constraint dynamics and counting physical degrees of freedom\label{sec3}}
The number of primary constraints as well as their corresponding Lagrange
multipliers are as follows:
\begin{center}
 \# $\phi^{\mu\nu}$ corresponding to $U_{\mu\nu}=6$\;\;\;\;\;\;\;\;\;\;\;\;\;\;\;\;\;\;\;\;\;\;\ \\
 \# $\Phi_i^{\mu\nu}$ corresponding to $\lambda^i_{\mu\nu} =2 \times
    6=12$\;\;\;\;\;\;\;\ \\
 \# $\Omega_{\mu\lambda\nu}$ corresponding to
    $V^{\mu\lambda\nu} =3 \times 6 =18$ \;\;\;\ \\
 \# total primary constraints $= 36$\;\;\;\;\;\;\;\;\;\;\;\;\;\;\;\;\;\;\;\;\;\
\end{center}
As in all constrained systems, the primary constraints should be
valid in the course of time. This means that their Poisson
brackets with the total Hamiltonian, which is responsible for the
dynamics of the system, should vanish. This process is the so
called "consistency of the constraints."\\
  It is important to remind the reader that at each step of consistency,
  two main things may happen. If a given constraint has
  non-vanishing Poisson bracket with some primary constraints, the
  corresponding Lagrange multiplier would be determined in terms
  of phase space variables. This is the case when the related
  constraint is second class \footnote {Remember that a set of
  constraints are second class if the matrix of their mutual
  Poisson brackets is non-singular. On the other hand, first
  class constraints have vanishing Poisson brackets with all of
  the constraints (at least on the constraint surface).}. TThe other possibility is that a new constraint
emerges as the consistency of the given constraint
and corresponding Lagrange multiplier is not determined.
These constraints at different levels of consistency are
called second level, third level, and so forth, which altogether
are remembered as secondary constraints. The process
of consistency will continue up to the last level in
which either a Lagrange multiplier is determined (when we
have a chain of second class constraints) or the consistency
is established identically (when the constraints in the corresponding
chain are first class). Now, we follow the consistency
procedure for our problem.

Consistency of $\Phi_i^{\mu\nu}$'s causes the following
 expressions vanish:
\begin{eqnarray}\label{7}
\chi_i^{\mu\nu}\equiv\{\Phi_i^{\mu\nu},H_T\}&=&-\frac{1}{2}
(\partial_i\Pi_0^{\mu\nu}-\frac{1}{2}\Pi_0^{\lambda\mu}
\xi_{\sigma\lambda}^\sigma\delta_i^\nu-\frac{1}{2}(g_{\lambda
i}A^{\mu\lambda\nu}-g_{\lambda\sigma}A^{\sigma\lambda\mu}\delta^\nu
_i)\nonumber\\&+&2\Pi_0^{\lambda\mu}\xi_{\lambda
i}^\nu)+\mu\longleftrightarrow\nu\ .
\end{eqnarray}
Since $\Phi_i^{\mu\nu}$ have vanishing Poisson brackets with all
primary constraints, no term containing Lagrange multipliers has
appeared in Eq. (\ref{7}). Therefore, consistency of 12 primary
constraints $\Phi_i^{\mu\nu}$ give 12 second level constraints
$\chi_i^{\mu\nu}$. consistency of $\chi_i^{\mu\nu}$ will be
investigated afterward.

For $\Omega_{\mu\lambda\nu}$ we have
\begin{eqnarray}\label{5}
\{\Omega_{\mu\lambda\nu},H_T\}&=&g_{\lambda\rho}\left (\
\xi^\rho_{\mu\nu}
-\frac{1}{2}(\xi^\sigma_{\mu\sigma}\delta^\rho_\nu
+\xi^\sigma_{\nu\sigma}\delta^\rho_\mu)\right )\ -\frac{1}{2}
(g_{\lambda\mu,i}
\delta^i_\nu+g_{\lambda\nu,i}\delta^i_\mu-g_{\mu\nu,i}\delta^i_\lambda)\nonumber\\
&-&\frac{1}{2}(U_{\lambda\mu}\delta^0_\nu+U_{\lambda\nu}\delta^0_\mu-U_{\mu\nu}\delta^0_\lambda).
\end{eqnarray}
The last term above includes Lagrange multipliers $U_{\mu \nu}$.
For $\mu=i$, $\nu=j$ and $\lambda=k$ this term vanishes and the
consistency of constraints $\Omega_{ikj}$ lead to the following second
level constraints:
\begin{eqnarray}\label{8}
\Theta_{ikj}=g_{k\rho}\left (\
\xi^\rho_{ij}-\frac{1}{2}(\xi^\sigma_{i\sigma}\delta^\rho_j
+\xi^\sigma_{j\sigma}\delta^\rho_i)\right )\
-\frac{1}{2}(g_{ki,j}+g_{kj,i}-g_{ij,k}).
\end{eqnarray}
The constraints $\Theta_{ikj}$ are in fact the same as
$\Psi_{ikj}$ given in Eq. (\ref{13}). We should investigate the
consistency of $\Theta_{ikj}$'s in the next level of consistency.
Let come back to Eq. (\ref{5}). The Lagrange multipliers
$U_{\alpha\beta}$ have appeared in the last term due to Poisson
brackets $\{\Omega_{\mu\lambda\nu}, \phi^{\alpha\beta}\}$ with one
of indices $\mu$ or $\lambda$ or $\nu$ considered as zero. This is
a ($12 \times 6$) rectangular matrix of rank 6. So it is possible
to divide $\Omega_{\mu\lambda\nu}$'s with at least one zero index
into two 6-member sets, as follows:
 \begin{eqnarray}\label{s1}
  \Omega^{(1)}=\left\{ \begin{array}{l} B_1\equiv \Omega_{001}\\ B_2\equiv
 \Omega_{002}\\ B_{11}\equiv \frac{1}{2}\Omega_{011}+\Omega_{101} \\ B_{22}\equiv
 \frac{1}{2}\Omega_{022}+\Omega_{202}\\ B_{12}\equiv
 \frac{1}{2}(\Omega_{012}+\Omega_{021})+\Omega_{102}\\ B'_{12}=\Omega_{012}-\Omega_{021}
 \end{array} \right. \hspace{1cm}
 \Omega^{(2)}=\left\{ \begin{array}{l} C_0\equiv \Omega_{000}\\ C_1\equiv
 \Omega_{010}\\ C_2\equiv \Omega_{020} \\ C_{11}\equiv \Omega_{011}-\frac{1}{2}\Omega_{101}\\
 C_{22}\equiv \Omega_{022}-\frac{1}{2}\Omega_{202}\\ C_{12}\equiv
 \frac{1}{2}(\Omega_{021}-\Omega_{012}-2\Omega_{102}) \end{array} \right.
 \end{eqnarray}
The constraints of the set $\Omega^{(1)}$ commute (i.e. has
vanishing Poisson brackets) with $\phi_{\alpha\beta}$'s and the
other set, $\Omega^{(2)}$ constitute a second class system with
$\phi^{\alpha\beta}$'s, so that the $6 \times 6$ matrix
$\{\Omega^{(2)}, \phi \}$ is nonsingular. We can redefine the
Lagrange multipliers $V_{\mu\lambda\nu}$ corresponding to the
division of the constraints $\Omega^{\mu\lambda\nu}$ into the
sets $\Omega^{ikj}$, $\Omega^{(1)}$ and $\Omega^{(2)}$, such that
 \begin{equation} \sum V_{\mu\lambda\nu}\Omega^{\mu\lambda\nu} = \sum
 V_{ikj}\Omega^{ikj} + \sum V_{(1)}\Omega^{(1)} + \sum
 V_{(2)}\Omega^{(2)}. \label{s20} \end{equation}
This gives
 \begin{eqnarray}\label{s2}
V_{(1)}=\left\{ \begin{array}{l} V^1_{(1)}\equiv V^{001}\\
V^2_{(1)}\equiv V^{002}\\ V^{11}_{(1)}\equiv
\frac{2}{5}(2V^{101}+V^{011}) \\ V^{22}_{(1)}\equiv
\frac{2}{5}(2V^{202}+V^{022})\\ V^{12}_{(1)}\equiv
 V^{012}+V^{021}\\ V'^{12}_{(1)}=V^{012}-\frac{1}{2}V^{102}
 \end{array} \right. \hspace{1cm}
 V_{(2)}=\left\{ \begin{array}{l} V^0_{(2)}\equiv V^{000}\\
V^1_{(2)}\equiv V^{010}\\V^2_{(2)}=V^{020}\\ V^{11}_{(2)}\equiv
\frac{2}{5}(2V^{011}-V^{101}) \\ V^{22}_{(2)}\equiv
\frac{2}{5}(2V^{022}-V^{202})\\ V^{12}_{(2)}\equiv
 V^{012}+V^{021}-V^{102}\end{array}. \right.
 \end{eqnarray}
 If we consider at this point the consistency of
$\phi^{\mu\nu}$'s, we get
\begin{eqnarray}\label{6}
\{\phi^{\mu\nu},H_T\}&=&-\frac{1}{4}m^2g_{\alpha\beta}g^{-1/2}
(\Pi_0^{\alpha\mu}\Pi_0^{\beta\nu}-3\Pi_0^{\alpha\beta}\Pi_0^{\mu\nu})
+\frac{1}{16}m^2g^{-1/2}g^{\mu\nu}\mathcal{G}_{\alpha\sigma\beta\lambda}
\Pi_0^{\alpha\beta}\Pi_0^{\sigma\lambda}\nonumber\\
&-&2m^2\Pi_0^{\mu\nu}-\sqrt{g}g^{\mu\nu}(\Lambda-3m^2)+\frac{1}{2}A^{\alpha\mu\beta}\left
(\ \xi^\nu_{\alpha\beta}
-\frac{1}{2}(\xi^\sigma_{\alpha\sigma}\delta^\nu_\beta+\xi^\sigma_{\beta\sigma}\delta^\nu_\alpha)\right )\
\nonumber\\
&-&\frac{1}{2}(A^{i\mu\nu}-\frac{1}{2}A^{\mu
i\nu})_{,i}+\frac{1}{2}(V^{0\mu\nu}-\frac{1}{2}V^{\mu 0
\nu})+\mu\longleftrightarrow\nu.
\end{eqnarray}
It can be seen easily that the last term in Eq. (\ref{6}) contains
Lagrange multipliers $V_{(2)}$ corresponding to the constraints in the
set $\Omega^{(2)}$ which have nonvanishing Poisson brackets with
$\phi^{\mu\nu}$'s. In this way the set of constraints
\begin{eqnarray}\label{17}
\phi \leftrightarrow \Omega^{(2)}
\end{eqnarray}
constitute a one-level 12-member family of second class
constraints which is constructed of two cross-conjugate chains
that determine 12 Lagrange multipliers $U_{\alpha\beta}$ and
$V_{(2)}$. In this description, we have used the language of ref.
\cite{refLoranshirzad} in classifying the families of constraints.
We just remind here that a family of constraints are determined as
a set of constraints which are resulted as the consistency of a
limited subset of primary constraints and make a close algebra of
Poisson brackets among themselves, and with the canonical Hamiltonian.
  Consistency of the set $\Omega^{(1)}$ gives 6 other constraints of
the next level as follows:
 \be \Psi^{(1)}=\left\{ \begin{array}{l} D_{11}\equiv \frac{1}{2}
 \Psi_{011}+\Psi_{101}\\ D_{22}\equiv \frac{1}{2}
 \Psi_{022}+\Psi_{202}\\ D_{12}\equiv
 \frac{1}{2}(\Psi_{012}+\Psi_{021})+\Psi_{102}
 \end{array} \right. \hspace{1cm}\Psi^{(2)}=\left\{ \begin{array}{l}
  D_j\equiv  \Psi_{0j0}~~~~~j=1,2
 \\ D'_{12}=\Psi_{012}-\Psi_{021}
 \end{array} \right. \label{s11} \ee
where $\Psi_{\alpha\lambda\beta}$ are Lagrangian constraints given in
Eq. (\ref{13}). We should continue to investigate consistency of
the above constraints in the next level. This will make the
meaning of the classification given in Eq. (\ref{s11}) more clear.
As can be seen, second level constraints $\Theta_{ikj}$,
$\Psi^{(1)}$, and $\Psi^{(2)}$ are 12 out of 18 Lagrangian
constraints (\ref{13}). The remaining 6 Lagrangian constraints
correspond to expressions (\ref{13}) including Lagrange
multipliers $U_{\mu\nu}$. In fact the equations of motion of
$g_{\mu\nu}$ gives $\dot{g}_{\mu\nu}=U_{\mu\nu}$. Putting this
into Eq. (\ref{5}) gives the corresponding Lagrangian constraints
(\ref{13}) for the cases which include time derivatives of the
metric.

Now we proceed to the next level by considering the consistency of
$\chi_i^{\mu\nu}$'s, $\Theta_{ikj}$'s and the sets $\Psi^{(1)}$
and $\Psi^{(2)}$. For $\Theta_{ikj}$'s we find
\begin{eqnarray}\label{10}
\{\Theta_{ikj},H_T\}=&&\frac{m^2}{4}[g_{k0}(\mathcal{G}_{\alpha
i\beta j}+\mathcal{G}_{\alpha j\beta
i})-\frac{1}{2}(g_{ki}(\mathcal{G}_{\alpha j\beta
0}+\mathcal{G}_{\alpha 0\beta j})\nonumber\\
+&&g_{kj}(\mathcal{G}_{\alpha i\beta 0}+\mathcal{G}_{\alpha 0\beta
i}))]\Pi_0^{\alpha\beta}
-g_{k0}(\xi^l_{ij,l}-\xi^\lambda_{i\sigma}\xi^\sigma_{j\lambda}
+\frac{1}{2}\xi^\lambda_{i\lambda}\xi^\sigma_{j\sigma})\nonumber\\
+&&\frac{1}{2}(g_{ki}(\xi^l_{j0,l}-\xi^\lambda_{j\sigma}\xi^\sigma_{0\lambda}
+\frac{1}{2}\xi^\lambda_{j\lambda}\xi^\sigma_{0\sigma})
+g_{kj}(\xi^l_{i0,l}-\xi^\lambda_{i\sigma}\xi^\sigma_{0\lambda}
+\frac{1}{2}\xi^\lambda_{i\lambda}\xi^\sigma_{0\sigma})) \nonumber\\
+&&4m^2(g_{k0}g_{ij}-\frac{1}{2}(g_{ki}g_{0j}+g_{kj}g_{0i}))\nonumber\\
+&&(\xi^\lambda_{ij}-\frac{1}{2}(\xi^\sigma_{j\sigma}\delta^\lambda_i
+\xi^\sigma_{i\sigma}\delta^\lambda_j))U_{k\lambda}-\frac{1}{2}(U_{kj,i}+U_{ki,j}-U_{ij,k})\nonumber\\
+&&g_{kl}\lambda^l_{ij}-\frac{1}{2}(g_{ki}\lambda^l_{jl}+g_{kj}\lambda^l_{il}).
\end{eqnarray}
Since $U_{\mu\nu}$ are determined previously, the Eq. (\ref{10})
should be considered as equations to find $\lambda^l_{mn}$. It is
easy to check that the matrix of coefficients of $\lambda^l_{mn}$'s
, i.e. $\{\Theta_{ikj}, \Phi_l^{mn}\}$, is nonsingular and
these Lagrange multipliers can be determined completely.\\
Consistency of constraints in the set $\Psi^{(2)}$ gives
 \begin{eqnarray}\label{D1}
 \{D_j, H_T\}&=&\frac{1}{8}m^2g^{-1/2}(g_{00}(\mathcal G_{0\mu
 j\nu}+\mathcal G_{j\mu 0\nu})-2g_{0j}\mathcal G_{0\mu 0\nu})
 \Pi_0^{\mu\nu}\nonumber\\&-&\frac{1}{2}\left (\
 g_{00}(\xi^i_{0j,i}-\xi^\lambda_{0\sigma}\xi^\sigma_{j\lambda}+
 \frac{1}{2}\xi^\sigma_{0\sigma}\xi^\lambda_{j\lambda})
 -2g_{0j}(\xi^i_{00,i}-\xi^\lambda_{0\sigma}\xi^\sigma_{0\lambda}
 +\frac{1}{2}\xi^\sigma_{0\sigma}\xi^\lambda_{0\lambda})\right )\
 \nonumber\\&-&\frac{1}{2}(g_{00}\lambda^i_{ji}+g_{0j}\lambda^i_{0i}
 -2g_{0i}\lambda^i_{0j}),~~~~~~~j=1,2
\end{eqnarray}
\begin{eqnarray}\label{D3}
\{D'_{12}, H_T\}&=&\frac{1}{8}m^2g^{-1/2}g_{10}(\mathcal G_{0\mu
2\nu}+\mathcal G_{2\mu 0\nu}) \Pi_0^{\mu\nu}-\frac{1}{2}
g_{10}(\xi^i_{02,i}-\xi^\lambda_{0\sigma}\xi^\sigma_{2\lambda}+
\frac{1}{2}\xi^\sigma_{0\sigma}\xi^\lambda_{2\lambda})\nonumber\\
&-&\frac{1}{2}(g_{10}\lambda^i_{2i}+g_{20}\lambda^i_{1i}-2g_{1i}
\lambda^i_{02})-1\longleftrightarrow
2
\end{eqnarray}
Since Lagrange multipliers $\lambda^i_{jk}$ are already
determined, Eqs. (\ref{D1})-(\ref{D3}) are three equations for
seven unknowns $V^1_{(1)}$, $V^2_{(1)}$, $V'^{12}_{(1)}$, and
$\lambda^i_{j0}$'s. Therefore, we should keep these equations in
mind and wait for finding four other equations which should be
solved together with (\ref{D1}) and (\ref{D3}) to find the above
unknowns. Anyhow, the consistency of $D_1$, $D_2$ and $D'_{12}$ do
not go further. Hence, we have a 6 member family of second class
constraints as
\be \begin{array}{lll}\label{fc}
 B_{1}\hspace{1cm}&B_{2}\hspace{1cm}&B'_{12} \\
 \downarrow\hspace{1cm}&\downarrow\hspace{1cm}&\downarrow\\
 D_{1}\hspace{1cm}&D_{2}\hspace{1cm}&D'_{12}
 \end{array} \ee
Consistency of $\chi_i^{\mu\nu}$'s leads the
following expression vanish:
\begin{eqnarray}\label{9}
\{\chi_i^{\mu\nu},H_T\}&=&(\Pi_0^{\lambda\mu}(\xi_{\lambda 0}^\nu
-
\xi^\sigma_{\lambda\sigma}\delta^\nu_0))_{,i}-\frac{1}{2}(g_{\lambda
0} A^{\mu\lambda\nu}-g_{\lambda\sigma}
A^{\sigma\lambda\mu}\delta_0^\nu)_{,i}\nonumber \;\;\;\;\;\;\;\;\;\;\;\;\;\;  \\
&-&\frac{1}{4}m^2g^{-1/2}((\mathcal{G}_{\alpha\lambda\beta i} +
\mathcal{G}_{\alpha
i\beta\lambda})\delta_0^\nu-\frac{1}{2}(\mathcal{G}_{\alpha\lambda\beta
0} + \mathcal{G}_{\alpha
0\beta\lambda})\delta_i^\nu)\Pi_0^{\lambda\mu}\Pi_0^{\alpha\beta}\nonumber\\
&-&4m^2(g_{\lambda i}\delta_0^\nu-\frac{1}{2}g_{\lambda 0}
\delta_i^\nu
)\Pi_0^{\lambda\mu}+(\Pi_0^{\sigma\lambda}\xi^\nu_{\sigma
0}+\Pi_0^{\sigma\nu}\xi^\lambda_{\sigma 0})\xi^\mu_{\lambda i}\nonumber\\
&-&\frac{1}{2}(\Pi_0^{\alpha\beta}\xi^\nu_{\beta
i}\delta^\mu_0+\Pi_0^{\alpha\mu}\xi^\nu_{0i})\xi^\rho_{\alpha\rho}
-\frac{1}{2}(\Pi_0^{\alpha\beta}\xi^\mu_{\alpha
0}
+\Pi_0^{\alpha\mu}\xi^\beta_{\alpha 0})\xi^\rho_{\beta\rho}\delta^\nu_i\nonumber\\
&+&(\xi^j_{\alpha
i,j}-\xi^\rho_{\alpha\sigma}\xi^\sigma_{i\rho}+\frac{1}{2}\xi^\rho_{\alpha\rho}
\xi^\sigma_{i\sigma})\Pi_0^{\alpha\mu}\delta^\nu_0\nonumber\\
&-&\frac{1}{4}(\Pi_0^{\sigma\alpha}\xi^\lambda_{\alpha\lambda}\delta_0^\mu
 +\Pi_0^{\sigma\mu}\xi^\lambda_{0\lambda})\xi^\rho_{\sigma\rho}\delta^\nu_i
 -\frac{1}{2}(\xi^j_{\lambda
 0,j}-\xi^\rho_{\lambda\sigma}\xi^\sigma_{0\rho}+\frac{1}{2}
 \xi^\rho_{\lambda\rho}\xi^\sigma_{0\sigma})\Pi_0^{\lambda\mu}\delta^\nu_i\nonumber\\
 &-&g_{\rho 0}(\xi^\nu_{\lambda i}-\frac{1}{2}\xi_{\lambda\sigma}^\sigma
 \delta^\nu_i)A^{\lambda\rho\mu}+g_{\rho\alpha}(A^{\alpha\rho\lambda}
 \xi^\nu_{\lambda i}\delta_0^\mu+A^{\alpha\rho\mu}\xi^\nu_{0i}\nonumber\\
 &-&\frac{1}{2}(A^{\alpha\rho\lambda}\xi^\sigma_{\lambda\sigma}\delta^\mu_0
 +A^{\alpha\rho\mu}\xi^\sigma_{0\sigma})\delta^\nu_i)
 -\Pi_0^{\sigma\mu}(\lambda^j_{\sigma i}\delta^\nu_j-\frac{1}{2}\lambda^j_{\sigma
 j}\delta^\nu_i)\nonumber\\&+&\frac{1}{2}(A^{\mu\sigma\nu}U_{i\sigma}
 -A^{\rho\sigma\mu}U_{\sigma\rho}\delta^\nu_i)+\frac{1}{2}(g_{\sigma
 i}V^{\mu\sigma\nu}-g_{\sigma\rho}V^{\rho\sigma\mu}\delta^\nu_i)+\mu\leftrightarrow\nu.
 \end{eqnarray}
Let us first consider 6 equations concerning the cases $\mu=j$ and
$\nu=k$ in Eq. (\ref{9}). It can be seen that the $6\times6$
matrix of coefficients of Lagrange multipliers $V^{ikj}$ is
nonsingular. Moreover, the corresponding equations do not include
the yet undetermined Lagrange multipliers $\lambda_{0j}^i$ and
$\lambda_{00}^i$. They include, however, the Lagrange multipliers
$U_{\mu\nu}$, $V_{(2)}$ and $\lambda^i_{lm}$ which are determined
previously. Therefore, the Eq. (\ref{9}) for the cases considered
can be used to determine 6 Lagrange multiplier $V^{ikj}$.
Nonsingularity of the ($6\times 6$) sub-matrix $\{\chi_i^{jk},
\Omega_{lmn}\}$
has an interesting meaning in the terminology of ref.
\cite{refLoranshirzad} on classifying the constraint families. To
this end, the set of constraints
\begin{eqnarray}\label{18}
\Omega_{ikj}\nwarrow\nearrow\Phi_i^{jk}\nonumber
\\\Theta_{ikj}\swarrow\searrow\chi_i^{jk},
\end{eqnarray}
constitute a family of 24-member, 2-level and cross-conjugate
second class system in which the consistency of constraints of the
first row give the constraints of the second row, while the
constraints at the end of any chain have nonvanishing Poisson
brackets with the constraints at the top of the other chain. We
can check that $\{\Omega_{ikj}, \Theta_{mln}\}$ as well as
$\{\Phi_i^{jk}, \chi_l^{mn}\}$ vanish.

Let us come back to Eq. (\ref{9}) and consider the case
($\mu=i,\nu=0$) or ($\mu=0,\nu=i$). We have four equations in this
case again for seven unknowns $V^1_{(1)}$, $V^2_{(1)}$,
$V'^{12}_{(1)}$ and $\lambda^i_{j0}$'s. These equations are in
fact, 4 equations which we were expecting, after Eq. (\ref{D3}).
Hence, we have 7 independent equations for 7 unknowns. In this way
the constraints $\chi_i^{0j}$ and their parents $\Phi_i^{0j}$
constitute an 8-member, 2-level family of second class constraints
shown as
 \begin{eqnarray}\label{19}
 \Phi_i^{0j}\nonumber\\ \downarrow \\ \chi_i^{0j}. \nonumber
 \end{eqnarray}

The only remaining case in Eq.(\ref{9}) is $\mu=\nu=0$. No term
containing $\lambda_{00}^i$ appears in Eq.(\ref{9}). This
corresponds to two constraints $\chi_i^{00}$ for which the term
including Lagrange multipliers $\lambda_i^{00}$ vanishes.
Consistency of $\chi_i^{00}$ lead to third level constraints
$\Sigma_i^{00}$ as
\begin{eqnarray}
\Sigma_i^{00}&=&-2(\Pi_0^{\lambda 0}\xi_{\lambda j}^j
)_{,i}-((g_{\lambda 0} A^{0\lambda 0}-g_{\lambda\sigma}
A^{\sigma\lambda 0}))_{,i}-8m^2g_{\lambda i} \Pi_0^{\lambda
0}+2(\Pi_0^{\sigma\lambda}\xi^0_{\sigma 0}+\Pi_0^{\sigma
0}\xi^\lambda_{\sigma 0})\xi^0_{\lambda
i}\nonumber\\&-&\frac{1}{2}m^2g^{-1/2}(\mathcal{G}_{\alpha\lambda\beta
i} + \mathcal{G}_{\alpha i\beta\lambda})\Pi_0^{\lambda
0}\Pi_0^{\alpha\beta}+2(\xi^j_{\alpha
i,j}-\xi^\rho_{\alpha\sigma}\xi^\sigma_{i\rho}+\frac{1}{2}
\xi^\rho_{\alpha\rho}\xi^\sigma_{i\sigma})\Pi_0^{\alpha
0}\nonumber\\
&-&(\Pi_0^{\alpha\beta}\xi^0_{\beta i}+\Pi_0^{\alpha
0}\xi^0_{0i})\xi^\rho_{\alpha\rho}-2(g_{\rho 0}\xi^0_{\lambda
i}A^{\lambda\rho 0}-g_{\rho\alpha}(A^{\alpha\rho\lambda}
 \xi^0_{\lambda i}+A^{\alpha\rho 0}\xi^0_{0i}))\nonumber\\&+&A^{0\lambda
0}U_{i\lambda}
 +g_{\sigma
 i}V^{0\sigma 0}.
\end{eqnarray}
Direct calculation shows that $\{\Sigma^{00}_i,\Phi_j^{00}\}$ is a
nonsingular matrix. Therefore, consistency of $\Sigma^{00}_i$
determines two Lagrange multipliers $\lambda_i^{00}$ and shows
that the constraints $\Sigma_i^{00}$ as well as their parents in
the corresponding chain are second class. In this way we have
derived a 6-member family of second class constraints gathered
in three-level chains as
 \begin{eqnarray}\label{22}
 \begin{array}{c}  \Phi_i^{00} \\ \downarrow \\ \chi_i^{00}\\ \downarrow
  \\ \Sigma_i^{00}
 \end{array}\ \ .\end{eqnarray}

Hence, 12 second class constraints in families (\ref{17})
determine 12 Lagrange multipliers $U_{\alpha\beta}$ and $V_{(2)}$ at
first level of consistency; 38 second class constraints in
families (\ref{18}), (\ref{19}) and (\ref{22}) determine 19
Lagrange multipliers $\lambda_i^{jk}$, $\lambda_i^{0j}$,
$V^{ikj}$, $V_{(1)}^1$, $V_{(1)}^2$ and ${V'}_{(1)}^{12}$ at
second level of consistency; and 6 second class constraints in
family (\ref{22}) determine 2 Lagrange multipliers
$\lambda_i^{00}$ at third level of consistency. We have a total of
S=12+38+6=56 second class constraints which determine 12+19+2=33
Lagrange multipliers. There remain 36-33=3 undetermined Lagrange
multipliers which are $V_{(1)}^{11}$, $V_{(1)}^{22}$, and
$V_{(1)}^{12}$ corresponding to primary constraints $B_{11},
B_{22}$ and $B_{12}$ given in (\ref{s1}).\\
 Let us recall that consistency of primary constraints $B_{11},
 B_{22}$, and $B_{12}$ gave us the second level constraints $D_{11},
 D_{22}$ and $D_{12}$ as in (\ref{s11}). Straightforward
 calculations shows that the Poisson brackets of $D_{11}, D_{22}$,
 and $D_{12}$ with the total Hamiltonian vanishes. Therefore, we
 collect the following constraints:
 \be\label{fc1}
 \begin{array}{lll}
 B_{11}\hspace{1cm}&B_{22}\hspace{1cm}&B_{12} \\
 \downarrow\hspace{1cm}&\downarrow\hspace{1cm}&\downarrow\\
 D_{11}\hspace{1cm}&D_{22}\hspace{1cm}&D_{12}
 \end{array}
  \ee
as a 6-member, 2-level and first class family of constraints. So
the number of first class constraints in the form of family
(\ref{fc1}) is $F=6$.

For the number of dynamical variables, using the famous formula
\cite{refHenoux}
\begin{eqnarray}\label{23}
D=N-S-2F,
\end{eqnarray}
where $N$ is the number of initially introduced variables in phase
space, we have $D=84-56-2\times 6=16$. This is in Hamiltonian
formalism. In Lagrangian formalism, we have half of this number as
dynamical variables. By dynamical variables, we mean those
variables which obey differential equations containing
accelerations. Taking a look on the complete action (\ref{4})
shows that the auxiliary variables $A^{\alpha\lambda\beta}$ are
not within these variables. Therefore, Eq.(\ref{23}) says that
after eliminating the redundant variables by using the constraints
and gauge fixing conditions, we have 8 dynamical equations for
$g_{\mu\nu}$'s and $\xi^\lambda_{\mu\nu}$'s. As we mentioned
before, the total number of variables $g_{\mu\nu}$ and
$\xi^\lambda_{\mu\nu}$ is 4 times greater than the number of
principle variable $g_{\mu\nu}$. Therefore, we conclude that the
number of dynamical variables is 8/4=2 out of six $g_{\mu\nu}$.

Notice that throughout our calculations concerning the constraint
structure of the NMG model we have kept the cosmological constant
term up to end. Hence, it is easy to find the constraint
structure of the model without cosmological model just by putting
$\Lambda=0$.\\
It should be noted that although the main structure of the
constraints and the number of degrees of freedom is the same, the
cosmological constant has a serious effect on the form of constrains
[see eq.(\ref{6})] as well as the final Hamiltonian, after
imposing the derived forms of Lagrange multipliers in the total
Hamiltonian (\ref{24}). Therefore, it can be expected that the
dynamics of the remaining physical degrees of freedom in the
reduced phase space is affected deeply by the presence of the
cosmological constant. Specially, particular solutions of the
equations of motion appear when $\Lambda\neq 0$ which are
forbidden in the absence of cosmological constant. However, our
purpose in this paper is not investigating the properties of
particular solutions, which stand beyond studying the constraint
structure of the system, although it could be interesting in turn.
\section{Concluding Remarks}
 In this paper we studied the Hamiltonian structure of new
 massive gravity model. This is a complicated model in 3d gravity
 that contains higher order derivatives as well as higher than
 quadratic terms. We used some form of the Palatini formalism in
 which combinations of Christoffel symbols, i.e., the variables
 $\xi^\lambda_{\mu\nu}$, are used as independent variables while
 their relation with derivatives of the metric is imposed as
 additional conditions in the Lagrangian using the auxiliary
 variables $A^{\alpha\lambda\beta}$.\\
  As is expected, the system is highly constrained with 36 primary
  and 26 secondary constraints, where 56 of them are second class
  and 6 are first class. This classification makes the system suitable to be
  studied more carefully in the context of constrained systems.
  For example, one may be interested in finding the generator of
  gauge transformations in terms of first class constraints and study more carefully the gauge
  symmetries of the system. Moreover, if one decides to fix the
  gauges, one needs to know carefully which gauge fixing conditions
  should be imposed. As is well-known \cite{refLoranshirzad} and \cite{refshirzad},
  for this proposes it is necessary to know the constraint
  structure of the system.\\
   We showed, finally in phase space there remain 16 physical
   variables. This means that there are 8 dynamical Lagrangian variables
   composed of the metric and Christoffel symbols. If one
   eliminates the Christoffel symbols, there remain 2
   dynamical degrees of freedom out of 6 components of the metric.
   This conclusion is in agreement with the result
   that the NMG model constitute 2 gravitons
   under linearization of the equations of motion.\\

 {\bf Acknowledgments: } the authors thank M. Alishahiha for his
useful comments and bringing their attention to NMG model.\\
\textbf{Note added}.-When completing our paper we observed reference
   \cite{refBlagojevic2} in which the total number of 2 degrees of
   freedom is derived in another approach.

 \end{document}